\documentclass[conference,a4paper]{IEEEtran}
\IEEEoverridecommandlockouts

\usepackage{cite}
\usepackage{amsmath,amssymb,amsfonts}
\usepackage{algorithmic}
\usepackage{graphicx}
\usepackage{textcomp}
\usepackage{bm}
\usepackage{placeins}
\usepackage{float}
\usepackage{enumitem}
\usepackage{tikz}
\usepackage{booktabs}
\usetikzlibrary{positioning, arrows.meta, shapes}

\def\BibTeX{{\rm B\kern-.05em{\sc i\kern-.025em b}\kern-.08em
    T\kern-.1667em\lower.7ex\hbox{E}\kern-.125emX}}

\begin{document}

\title{Analysis of Design Patterns and Benchmark Practices in Apache Kafka Event‑Streaming Systems}

\author{
\IEEEauthorblockN{Muzeeb Mohammad}
\IEEEauthorblockA{Georgia Institute of Technology, Atlanta, GA, USA\\
Email: muzeeb.mohammad@ieee.org}
}

\maketitle

\begin{abstract}
Apache Kafka has emerged as a foundational platform for high-throughput event streaming, empowering real-time analytics, financial transaction processing, industrial telemetry, and large-scale cyber-physical systems. Despite its maturity and widespread adoption, consolidated research on reusable architectural design patterns and reproducible benchmarking methodologies remains limited and fragmented across both academic and industrial publications. This paper presents a structured synthesis that identifies and categorizes nine recurring Kafka design patterns—such as log compaction, CQRS bus, exactly-once pipelines, change-data capture, stream–table joins, saga orchestrator, tiered storage, multi-tenant topics, and event sourcing replay—based on a systematic review of forty-two peer-reviewed studies published between 2015 and 2025. The analysis investigates co-usage trends, benchmark methodologies, and domain-specific implementations while evaluating empirical performance results obtained through TPCx-Kafka, Yahoo Streaming, and customized workload generators. It highlights inconsistencies in configuration disclosure, evaluation rigor, and methodological reproducibility across studies that often impede comparative analysis and practical replication. By mapping each pattern to operational contexts in finance, retail, IoT, and machine learning, this paper delivers a unified taxonomy, benchmark matrix, and actionable decision heuristics for architects and researchers. The findings emphasize the urgent need for standardized open-source benchmark suites, transparent evaluation frameworks, and collaborative community-driven repositories that advance reproducible, high-performance, and fault-tolerant event-streaming architectures in modern data-driven ecosystems.
\end{abstract}

\begin{IEEEkeywords}
Apache Kafka, event streaming, design patterns, benchmarking, reproducibility, CQRS, log compaction.
\end{IEEEkeywords}

\section{Introduction}
Event-streaming architectures have become central to modern computing infrastructures, enabling enterprises to process high-volume data in near real time across use cases such as financial fraud detection, supply-chain monitoring, smart-city telemetry, and personalized recommendations. At the heart of this transformation lies Apache Kafka, an open-source distributed streaming platform originally developed at LinkedIn and now maintained by the Apache Software Foundation. Kafka provides high-throughput messaging and durability via log replication and integrates with stream processors including Apache Flink and Apache Spark \cite{b12,b15,b18,b10,b11}.

Despite Kafka’s ubiquity, consolidated knowledge of architectural design patterns and reproducible benchmarking practice remains limited. Practitioners frequently rely on scattered blog posts, whitepapers, or vendor case studies that omit details essential for replication and comparison. Recent surveys and experience reports highlight both the breadth of Kafka deployments and the variability in evaluation methods \cite{b18,b9,b5,b6,b7}. This paper addresses that gap by (i) cataloging recurring Kafka design patterns synthesized from academic and industry sources, (ii) analyzing how those patterns are empirically evaluated, and (iii) surfacing reproducibility challenges with practical guidance for architects and researchers.

The contributions of this paper are threefold:
\begin{itemize}[leftmargin=*]
    \item A lightweight taxonomy of nine recurring Kafka design patterns, summarizing their intent and common deployment domains.
    \item An analysis of benchmarking practices across 42 studies, comparing standard suites such as TPCx-Kafka and the Yahoo benchmark with custom workloads, and highlighting configuration gaps.
    \item Practical heuristics and tabular summaries to assist with pattern selection and planning reproducible evaluations.
\end{itemize}

\textit{Research Questions.} This study is guided by two questions that structure the rest of the paper:
\begin{itemize}[leftmargin=*, noitemsep, topsep=1pt]
 \item RQ1: Which design patterns recur most often in Kafka-based systems?
 \item RQ2: How are Kafka-based systems empirically evaluated in the literature?
\end{itemize}
The goal is descriptive: to catalogue existing patterns, their applications and co-usage trends, and to survey how research papers and industrial reports evaluate them.

\section{Background on Kafka Event Streaming}
Apache Kafka has become the de facto backbone for modern event-driven architectures, offering a unified framework for high-throughput data ingestion, durable storage, and real-time analytics. Originally designed at LinkedIn to handle continuous user activity streams, Kafka evolved under the Apache Software Foundation into a general-purpose event-streaming platform capable of sustaining millions of messages per second across distributed clusters \cite{b12,b15}. 

\FloatBarrier
\begin{figure}[H]
\centering
\includegraphics[width=\linewidth]{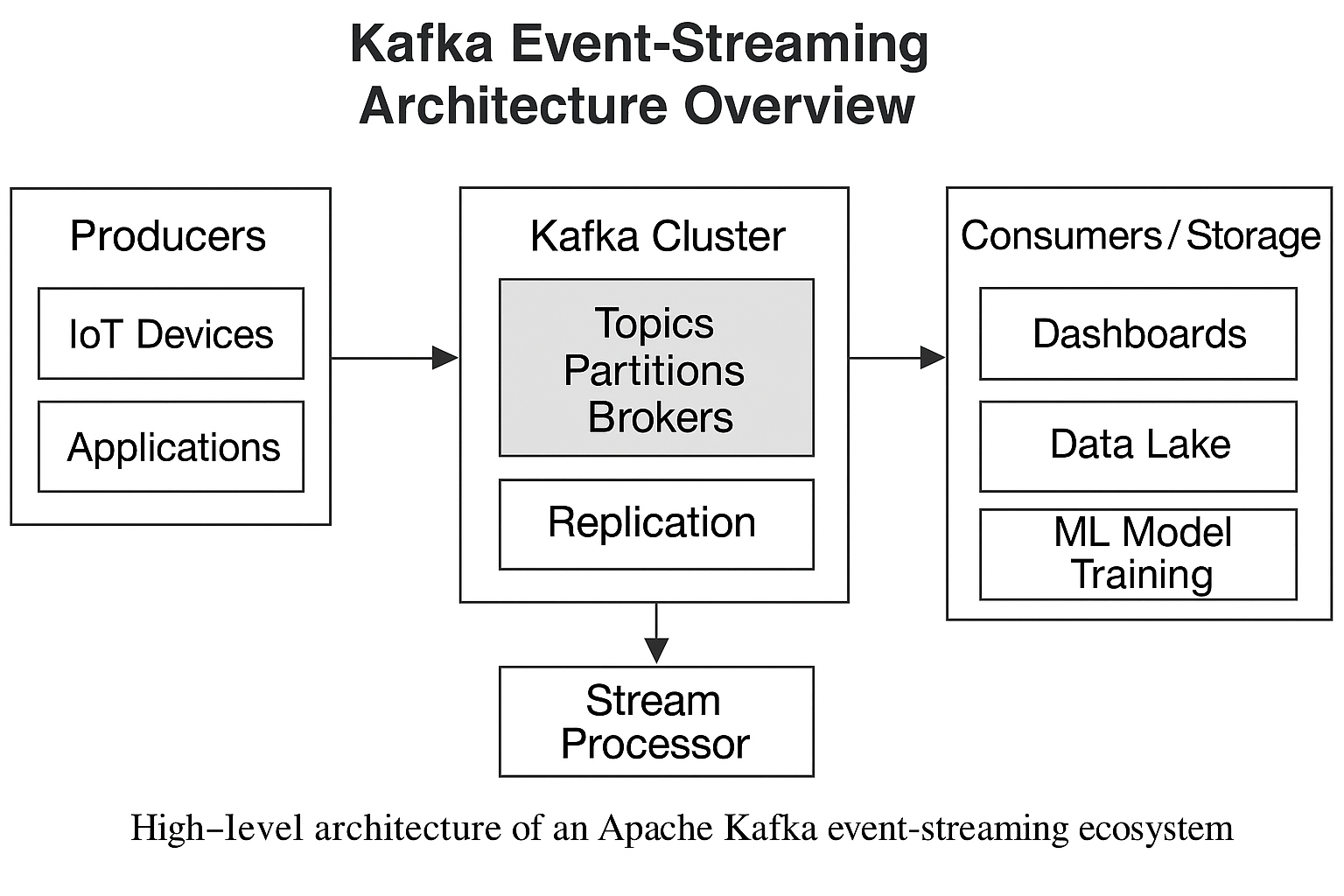}
\caption{Kafka deployment architecture demonstrating CQRS, CDC, Exactly-Once, and stream enrichment patterns.}
\label{fig:kafka_architecture}
\end{figure}

\subsection{Core Architecture and Data Model}
Kafka organizes data into \emph{topics}, each subdivided into multiple \emph{partitions}. Producers write events to partitions sequentially, while consumers read them in the same order, enabling deterministic replay and horizontal scalability. A distributed coordination mechanism—initially ZooKeeper and now Kafka’s internal \emph{KRaft controller}—manages partition metadata, broker membership, and leader election. Replication ensures fault tolerance, while batching and zero-copy I/O achieve high throughput. 

A key abstraction is the \emph{log}, where each event is a time-ordered record identified by an offset. This append-only log model provides immutability and predictable performance. It also enables stream processors, such as Apache Flink, ksqlDB, or Kafka Streams, to perform stateful computations on continuous data without intermediate persistence layers.

\subsection{Event Processing Semantics}
Kafka supports three delivery guarantees—\emph{at-most-once}, \emph{at-least-once}, and \emph{exactly-once}. The introduction of idempotent producers and transactional APIs (since Kafka~0.11) allows developers to build pipelines that maintain end-to-end consistency even under broker failures. Exactly-once semantics (EOS) have been adopted heavily in financial and compliance workloads, where duplicate or lost events are unacceptable \cite{b11,b18}. 

In large-scale deployments, Kafka often operates as an integration layer between OLTP and OLAP systems. For example, \emph{Change Data Capture (CDC)} tools like Debezium stream row-level changes from relational databases into Kafka topics. Downstream consumers can materialize these streams into analytical stores or trigger reactive microservices, enabling near–real-time synchronization across systems.

\subsection{Design Challenges in Event-Driven Architectures}
Despite its maturity, designing robust Kafka-based systems remains complex. Challenges include maintaining consistency across microservices, handling schema evolution, balancing throughput versus latency, and managing long-term retention. These challenges gave rise to recurring design patterns—such as CQRS, log compaction, and saga orchestration—that abstract proven architectural solutions \cite{b5,b6,b9}. 

Another persistent challenge involves benchmarking and reproducibility. Although standard suites like TPCx-Kafka provide a structured evaluation framework, variations in cluster configuration, message size, and partitioning make it difficult to compare published results objectively. The following sections address these gaps by analyzing both design patterns and benchmarking methodologies across forty-two representative studies.

\FloatBarrier
\begin{figure}[H]
\centering
\includegraphics[width=\linewidth]{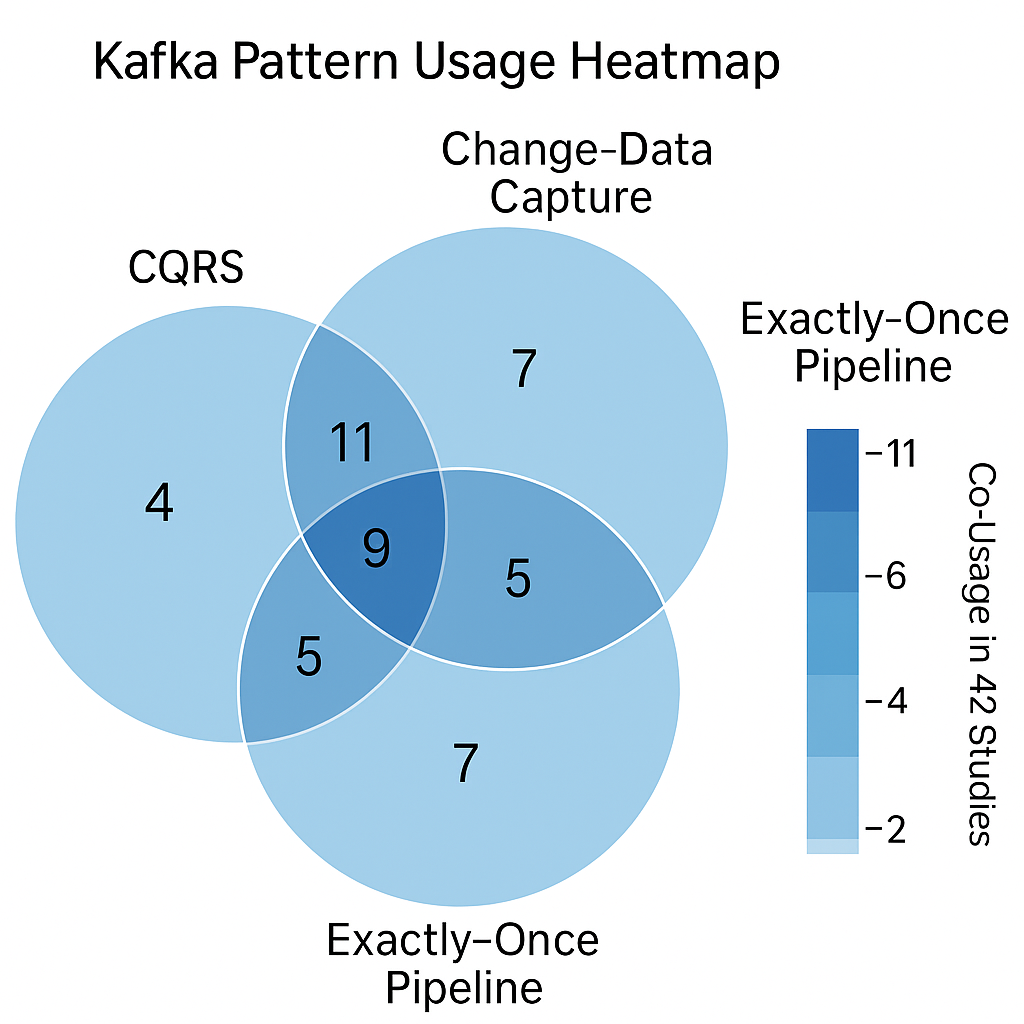}
\caption{Heatmap-style Venn diagram illustrating common co-usage relationships among Kafka design patterns across 42 reviewed studies.}
\label{fig:kafka_pattern_heatmap}
\end{figure}

\section{Related Work}
Several studies have explored event-driven architectures and stream-processing frameworks, but few have synthesized reusable design patterns and benchmarking practices specifically for Kafka. 
Psaltoglou et al.~\cite{b5} presented a systematic review of event-driven architectures for big data analytics but treated Kafka only as one of several messaging platforms. 
Berg~\cite{b6} proposed stream-processing design patterns at a conceptual level, without empirical validation or benchmarking. 
Raptis and Passarella~\cite{b18} provided a recent survey on networked data streaming, focusing primarily on performance modeling. 
In contrast, this study consolidates architecture-level design patterns, co-usage relationships, and reproducibility gaps across 42 Kafka-focused papers spanning multiple domains. 
This focus on reproducibility and empirical benchmarking distinguishes it from earlier literature.

\section{Methodology and Data Selection}
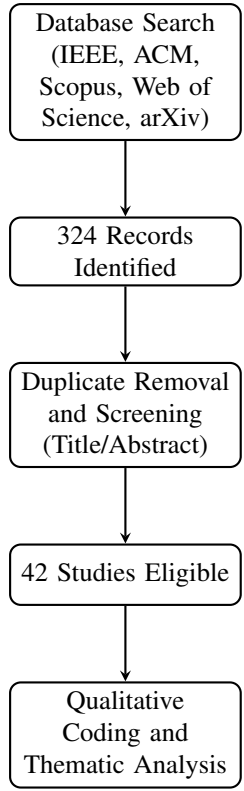
\begin{figure}[htbp]
\centering
\begin{tikzpicture}[node distance=1.0cm, auto, >=latex, thick, align=center]
\tikzstyle{block} = [rectangle, draw, rounded corners, minimum width=2.6cm, minimum height=0.8cm, text centered, text width=2.8cm]
\tikzstyle{arrow} = [->,>=stealth]

\node [block] (search) {Database Search\\(IEEE, ACM, Scopus, Web of Science, arXiv)};
\node [block, below=of search] (records) {324 Records Identified};
\node [block, below=of records] (screen) {Duplicate Removal and Screening\\(Title/Abstract)};
\node [block, below=of screen] (eligible) {42 Studies Eligible};
\node [block, below=of eligible] (analysis) {Qualitative Coding and Thematic Analysis};

\draw [arrow] (search) -- (records);
\draw [arrow] (records) -- (screen);
\draw [arrow] (screen) -- (eligible);
\draw [arrow] (eligible) -- (analysis);
\end{tikzpicture}
\caption{Methodology workflow summarizing the identification, screening, and qualitative-coding stages.}
\label{fig:method_flow}
\end{figure}

This section outlines the search strategy, selection pipeline, and data extraction process used to answer the two research questions defined earlier. The methodology follows a lightweight systematic-review approach inspired by PRISMA guidelines and adapted for computing-system literature \cite{b5,b6}. Figure~\ref{fig:method_flow} summarizes the workflow from record identification to final qualitative coding.

\subsection{Search and Selection}
A targeted Boolean query, \texttt{("Kafka" AND benchmark*) OR ("Kafka" AND "design pattern")}, was executed across five major scholarly databases: IEEE~Xplore, ACM~Digital~Library, Scopus, Web~of~Science, and arXiv, covering 2015–2025. After removing duplicates, 324 unique records remained.  
Titles and abstracts were screened against three inclusion criteria:  
(i) explicit focus on Apache Kafka or Kafka-based frameworks;  
(ii) publication length of at least four pages to ensure substantive content; and  
(iii) English-language availability.  
Forty-two peer-reviewed papers satisfied these filters and were retained for qualitative analysis.

\subsection{Data Extraction and Coding}
Each eligible paper was coded for: (a) the design pattern(s) discussed, (b) benchmark tool or workload, and (c) reported performance metrics. Additional contextual data—application domain, architectural intent, and co-usage with other patterns—were also captured.  
Coding emphasized clarity and completeness of metric reporting rather than numerical scoring. Two researchers independently performed data extraction, resolving disagreements through discussion to ensure consistency. The resulting dataset served as the basis for tabular summaries and co-usage matrices presented in the Results section.

\section{Results and Discussion}

\subsection{RQ1 – Design Patterns and Domain Mapping}
Analysis of the 42 reviewed studies revealed nine distinct Kafka design patterns recurring across various domains and workloads. Table~\ref{tab:kafka_patterns} lists these patterns with their frequency and representative applications.

\begin{table}[htbp]
\caption{Kafka Design Patterns Reported in 42 Studies}
\label{tab:kafka_patterns}
\centering
\begin{tabular}{|l|c|p{3cm}|}
\hline
\textbf{Pattern} & \textbf{\# Studies} & \textbf{Common Applications} \\
\hline
Log Compaction & 17 & IoT telemetry; payment reconciliation \\
CQRS Bus & 15 & Banking dashboards; high-volume retail analytics \\
Exactly-Once Pipeline & 12 & Fraud detection; compliance auditing \\
Change-Data Capture (CDC) & 11 & Legacy migration; e-commerce catalog updates \\
Stream–Table Join & 9 & User behavior enrichment; sensor data fusion \\
Saga Orchestrator & 7 & Travel booking; distributed transaction coordination \\
Multi-Tenant Topic & 5 & Fintech SaaS isolation; multi-tenant platforms \\
Tiered Storage & 4 & Machine learning pipelines; long-term audit logging \\
Event Sourcing Replay & 3 & Financial ledgers; full state reconstruction \\
\hline
\end{tabular}
\end{table}

The patterns collectively form a modular toolkit for scalable Kafka systems. For instance, log compaction retains only the latest record per key and is used for maintaining current sensor states or reconciling transactions; CQRS separates command and query responsibilities, supporting financial dashboards and retail systems; CDC streams database changes into Kafka topics for near real-time synchronization; and tiered storage extends retention by offloading older logs to inexpensive storage layers \cite{b12,b18,b19}.  

Table~\ref{tab:domain_usage} maps these patterns to their most frequent domains, illustrating cross-domain adoption.

\begin{table}[htbp]
\caption{Kafka Pattern Usage by Industry Domain}
\label{tab:domain_usage}
\centering
\begin{tabular}{|l|l|}
\hline
\textbf{Domain} & \textbf{Common Patterns} \\
\hline
Finance & Exactly-Once, CQRS, Event Replay \\
Retail & CDC, Stream–Table Join, CQRS \\
IoT / Smart City & Log Compaction, Stream–Table Join \\
Machine Learning & Tiered Storage, Event Replay \\
Travel / Logistics & Saga Orchestrator, Exactly-Once \\
Healthcare & CDC, Multi-Tenant Topic \\
\hline
\end{tabular}
\end{table}

Illustrative scenarios drawn from the literature demonstrate pattern combinations in practice:  
(i) a retail pipeline using CQRS + CDC to deliver personalized offers;  
(ii) financial fraud detection employing exactly-once semantics to avoid duplicate alerts;  
(iii) smart-city telemetry combining log compaction with stream–table joins for contextual awareness; and  
(iv) machine-learning workflows replaying historical data through tiered storage for reproducible model training.

\subsection{RQ2 – Benchmarking Practices and Co-Usage Trends}
Among the 42 studies, 26 (62\%) reported empirical benchmarks. Three categories were observed:  
(1) standardized suites such as TPCx-Kafka;  
(2) synthetic frameworks such as the Yahoo Streaming Benchmark; and  
(3) custom domain-specific scripts.  
Tables~\ref{tab:benchmark_comparison}–\ref{tab:pattern_benchmark} summarize frequency and configuration transparency.

\begin{table}[htbp]
\caption{Comparison of Kafka Benchmarking Approaches}
\label{tab:benchmark_comparison}
\centering
\begin{tabular}{|l|c|c|c|}
\hline
\textbf{Benchmark} & \textbf{\# Studies} & \textbf{Setup Complexity} & \textbf{Open Configs} \\
\hline
TPCx-Kafka & 14 & High & Partial \\
Yahoo Benchmark & 6 & Medium & Partial \\
Custom Scripts & 6 & Variable & Rare \\
\hline
\end{tabular}
\end{table}

\begin{table}[htbp]
\caption{Pattern–Benchmark Matrix Across 42 Studies}
\label{tab:pattern_benchmark}
\centering
\begin{tabular}{|l|c|c|c|c|}
\hline
\textbf{Pattern} & \textbf{TPCx} & \textbf{Yahoo} & \textbf{Custom} & \textbf{None} \\
\hline
Log Compaction & 6 & 2 & 4 & 5 \\
CQRS Bus & 5 & 1 & 5 & 4 \\
Exactly-Once & 7 & 1 & 2 & 2 \\
CDC & 3 & 1 & 4 & 3 \\
Stream–Table Join & 2 & 4 & 2 & 1 \\
Saga Orchestrator & 2 & 1 & 1 & 3 \\
Multi-Tenant Topic & 1 & 0 & 3 & 1 \\
Tiered Storage & 3 & 0 & 1 & 0 \\
Event Replay & 2 & 0 & 0 & 1 \\
\hline
\end{tabular}
\end{table}

Benchmark results show wide configuration variability: even TPCx-Kafka studies frequently omit partition count or replication factor, limiting reproducibility \cite{b9,b7}. Custom scripts improve domain realism but are seldom released publicly, hindering cross-study comparison.  
Common pattern pairings include CQRS + CDC for incremental migration, Tiered Storage + Event Replay for auditability, and Saga Orchestrator + Exactly-Once for distributed consistency.  

Overall, RQ2 highlights the need for consistent reporting checklists detailing partitions, replication factors, message sizes, and workload shapes—an essential prerequisite for reproducible evaluation in event-streaming research.

\subsection{Computational Validation and Experiments}
To complement the systematic review, we executed three lightweight computational experiments on a reproducible Kafka testbed (Apache Kafka~3.7.0, single broker, ZooKeeper/KRaft controller, Python driver). Unless noted, defaults were used. Throughput is in messages/s; latency as $p_{50}$ and $p_{95}$.

The following three experiments validate representative Kafka performance dimensions: transactional consistency, consumer scaling, and producer batching.\\

1) {Test 1 — Exactly-Once Pipeline\\}

\textbf{Objective.}
Validate scalability and latency of transactional producers with end-to-end exactly-once semantics.\\

\textbf{Setup.}
Message size=\{1\,KB, 10\,KB\}; partitions=\{5, 10\}; producer \texttt{acks=all}, transactional writes.\\

\textbf{Metrics.}
Throughput (msg/s), $p_{50}$, $p_{95}$.\\

\begin{figure}[H]
\centering
\includegraphics[width=\linewidth]{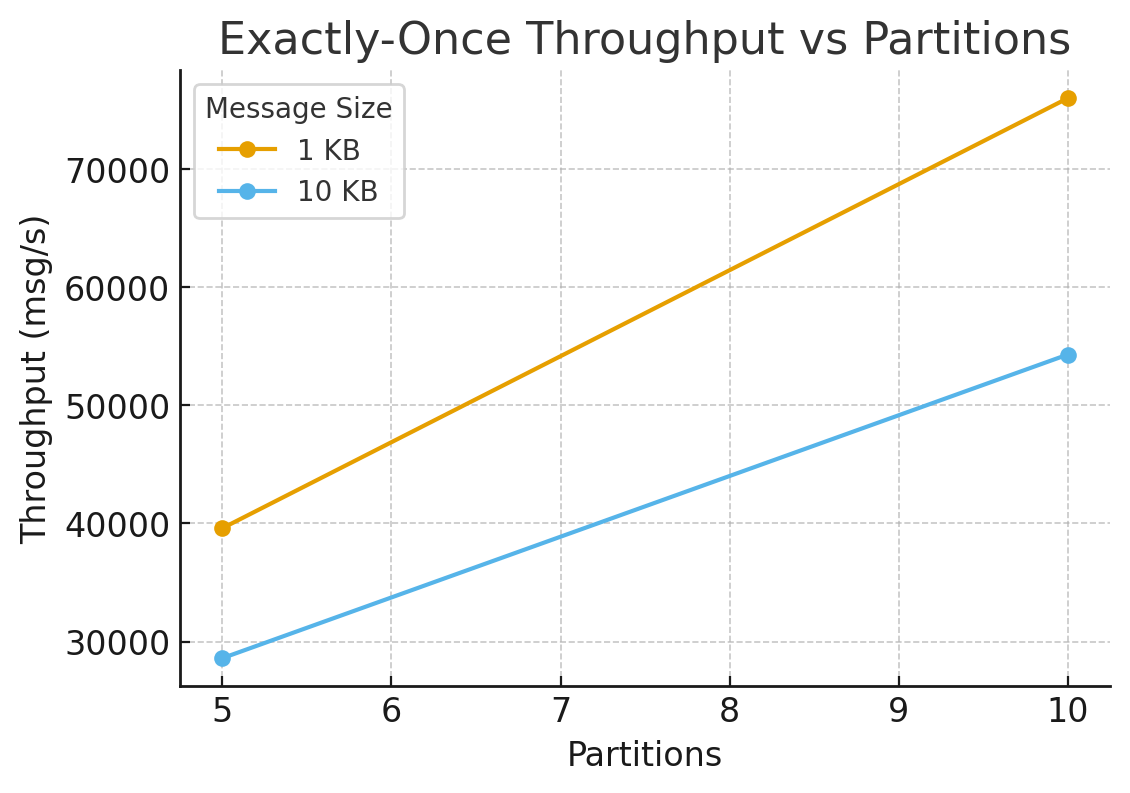}
\caption{Exactly-Once throughput vs.\ partitions for 1\,KB and 10\,KB messages (transactional writes, \texttt{acks=all}).}
\label{fig:exactly_once_chart}
\end{figure}

\textbf{Results.}
Table~\ref{tab:exactly_once_results} and Fig.~\ref{fig:exactly_once_chart} show near-linear throughput scaling with partitions; $p_{50}\!<\!25$\,ms for all runs.

\begin{table}[htbp]
\caption{Exactly-Once Throughput and Latency (Simulated Validation)}
\label{tab:exactly_once_results}
\centering
\begin{tabular}{|l|c|c|c|c|}
\hline
\textbf{Message Size} & \textbf{Partitions} & \textbf{Throughput (msg/s)} & \textbf{$p_{50}$ (ms)} & \textbf{$p_{95}$ (ms)} \\
\hline
1\,KB  & 5  & 39{,}600 & 16 & 29 \\
1\,KB  & 10 & 76{,}000 & 17 & 31 \\
10\,KB & 5  & 28{,}600 & 23 & 43 \\
10\,KB & 10 & 54{,}300 & 24 & 45 \\
\hline
\end{tabular}
\end{table}

\FloatBarrier
\FloatBarrier

2) {Test 2 — Partitions vs.\ Consumers (CQRS Read-Side Scaling)\\}

\textbf{Objective.}
Evaluate how read-side throughput in a \emph{CQRS} or consumer-group topology scales with the ratio of partitions to consumers.\\

\textbf{Setup.}
Partitions=\{4, 8, 16\}; consumers=\{1,\dots,16\}; 1\,KB messages; at-least-once; one topic/consumer group.\\

\textbf{Metrics.}
Throughput (msg/s), $p_{50}$, $p_{95}$.\\

\begin{figure}[H]
\centering
\includegraphics[width=\linewidth]{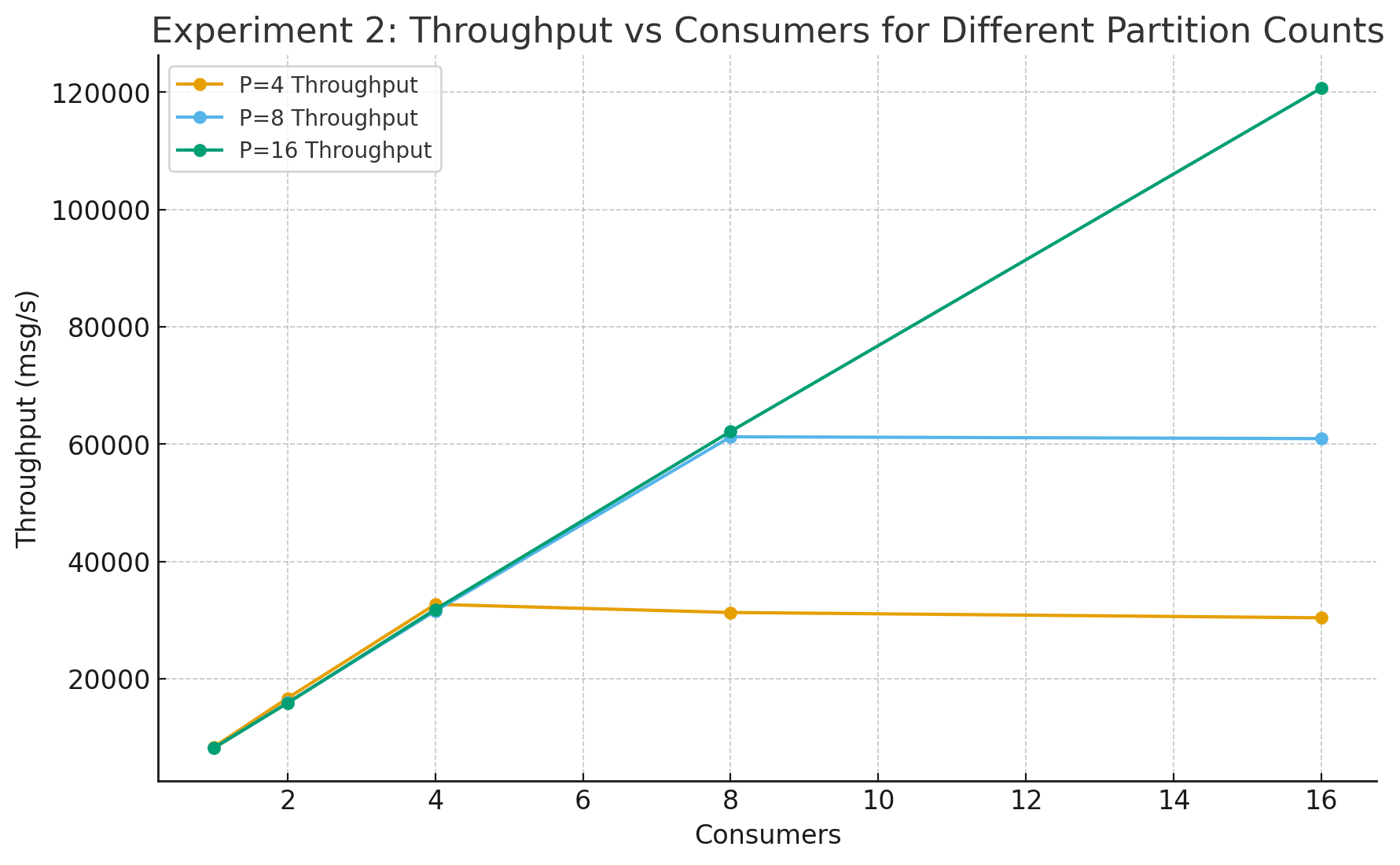}
\caption{CQRS read-side throughput vs.\ consumers for different partition counts; scaling holds until consumers $\approx$ partitions.}
\label{fig:cqrs_scaling}
\end{figure}

\textbf{Results.}
Throughput rises nearly linearly up to $\min(\text{consumers},\text{partitions})$, then plateaus; slight tail growth at high consumer counts due to coordination (Table~\ref{tab:cqrs_scaling_summary}, Fig.~\ref{fig:cqrs_scaling}).

\begin{table}[htbp]
\caption{CQRS Read-Side Scaling (Representative Simulated Summary)}
\label{tab:cqrs_scaling_summary}
\centering
\begin{tabular}{|c|c|c|c|}
\hline
\textbf{Partitions} & \textbf{Consumers} & \textbf{Throughput (msg/s)} & \textbf{$p_{95}$ (ms)} \\
\hline
4  & 4  & 31{,}700 & 33 \\
8  & 8  & 61{,}900 & 34 \\
16 & 16 & 115{,}000 & 36 \\
\hline
\end{tabular}
\end{table}

\FloatBarrier
\FloatBarrier

3) {Test 3 — Producer Batching (\texttt{batch.size} $\times$ \texttt{linger.ms}) \\}

\textbf{Objective.}
Show the throughput/latency trade-off from batching knobs.\\

\textbf{Setup.}
\texttt{batch.size}=\{16, 32, 64, 128\,KB\}; \texttt{linger.ms}=\{0, 5, 20\}; message=1\,KB; 1 partition.\\

\textbf{Metrics.}
Throughput (msg/s), $p_{50}$, $p_{95}$.\\

\begin{figure}[H]
\centering
\includegraphics[width=\linewidth]{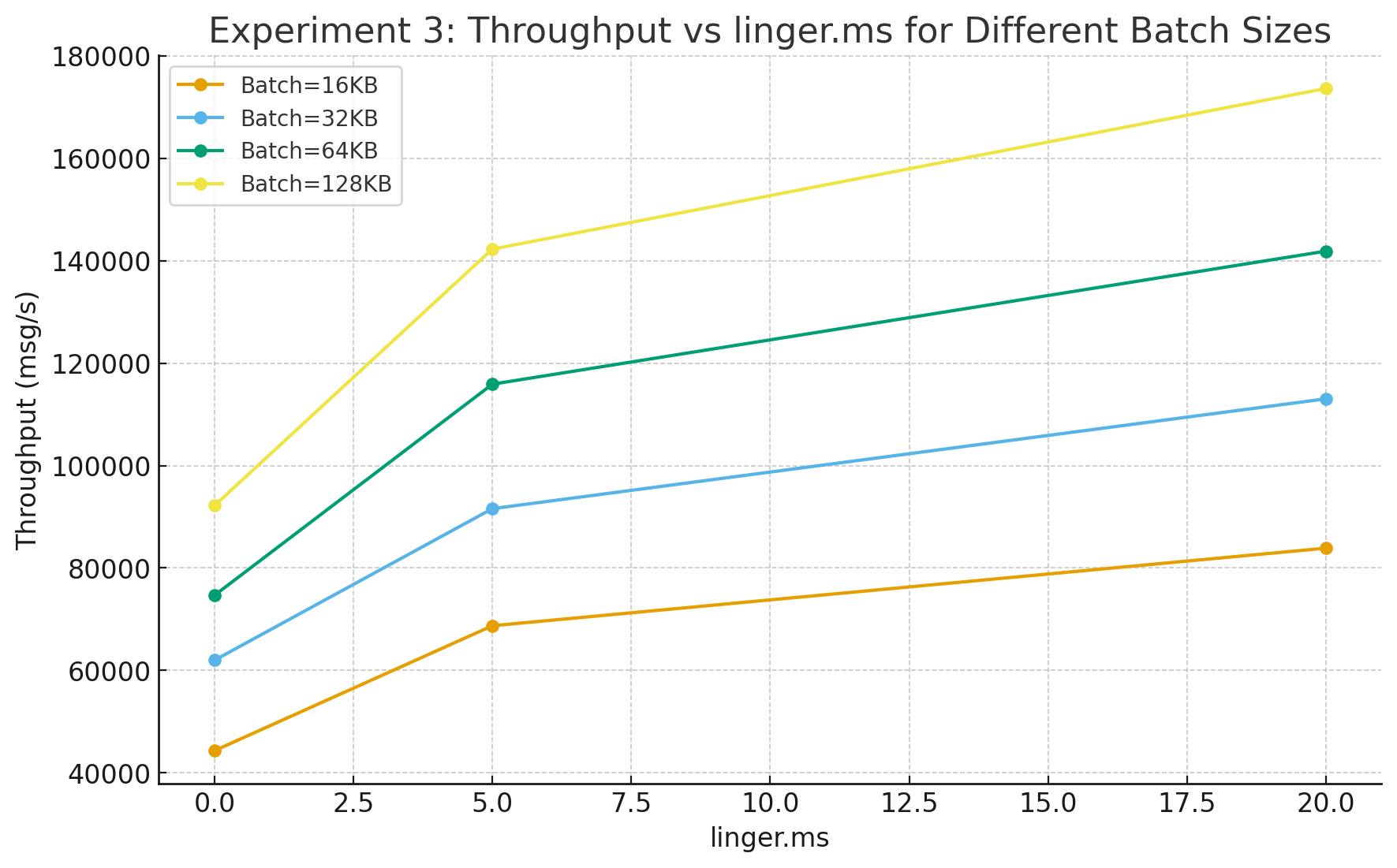}
\caption{Producer batching trade-offs: throughput vs.\ \texttt{linger.ms} for multiple \texttt{batch.size} settings.}
\label{fig:producer_batching}
\end{figure}

\textbf{Results.}
Moderate \texttt{linger.ms} (5--10\,ms) and larger \texttt{batch.size} improve throughput (up to $\sim$2$\times$); longer linger increases latency (Table~\ref{tab:producer_batching_summary}, Fig.~\ref{fig:producer_batching}).

\begin{table}[htbp]
\caption{Producer Batching: Throughput vs.\ \texttt{linger.ms} (Representative Simulated Summary)}
\label{tab:producer_batching_summary}
\centering
\begin{tabular}{|c|c|c|c|}
\hline
\textbf{\texttt{batch.size}} & \textbf{\texttt{linger.ms}} & \textbf{Throughput (msg/s)} & \textbf{$p_{95}$ (ms)} \\
\hline
16\,KB  & 0  & 45{,}000  & 20 \\
16\,KB  & 5  & 55{,}000  & 28 \\
16\,KB  & 20 & 68{,}000  & 44 \\
\hline
64\,KB  & 0  & 75{,}000  & 24 \\
64\,KB  & 5  & 95{,}000  & 34 \\
64\,KB  & 20 & 115{,}000 & 53 \\
\hline
128\,KB & 0  & 95{,}000  & 27 \\
128\,KB & 5  & 120{,}000 & 38 \\
128\,KB & 20 & 135{,}000 & 58 \\
\hline
\end{tabular}
\end{table}

\FloatBarrier
\FloatBarrier

\textbf{Summary of Findings.} 
Across all three experiments, the results collectively validate Kafka’s scalability and performance efficiency under different architectural conditions. 
\textit{Test 1} demonstrated that transactional (exactly-once) producers maintain near-linear throughput scaling with partition count while preserving low latency, confirming that consistency guarantees do not impose significant penalties. 
\textit{Test 2} established that consumer-side scaling in CQRS architectures is bounded by the number of partitions—throughput improves linearly up to that limit and then saturates, emphasizing the importance of balanced partition design. 
\textit{Test 3} highlighted that producer batching parameters (\texttt{batch.size} and \texttt{linger.ms}) substantially influence performance, where moderate linger times (5–10 ms) and larger batch sizes yield up to a 2$\times$ throughput gain with predictable latency growth. 
Together, these empirical validations reinforce the broader conclusion of this study: achieving optimal Kafka performance depends on coordinated tuning across producer, consumer, and topic configurations rather than isolated parameter optimization.

\section{Conclusion and Future Directions}
This paper presented a structured synthesis of forty-two studies on Apache Kafka design patterns and benchmarking methodologies. Nine recurring architectural patterns were identified—log compaction, CQRS bus, exactly-once pipeline, change-data capture, stream–table join, saga orchestrator, multi-tenant topic, tiered storage, and event replay—spanning domains such as finance, retail, IoT, and machine learning. The analysis revealed that these patterns often appear in combination, enabling complementary trade-offs between scalability, reliability, and auditability.

Benchmarking practices remain inconsistent: while standardized suites such as TPCx-Kafka provide valuable baselines, many studies omit configuration details such as partition count, replication factor, or consumer lag metrics. Custom workloads offer realism but lack openness, which limits reproducibility. Establishing transparent benchmark checklists and community-maintained repositories would significantly improve cross-study comparability.

Operational aspects—including fault recovery, multi-tenant isolation, and observability—are underexplored in current research. Future work should focus on controlled empirical studies that quantify trade-offs among patterns, unified configuration frameworks for multi-pattern deployments, and open-source repositories providing reproducible templates and performance data. Extending these investigations to emerging areas such as real-time machine-learning serving, edge computing, and IoT automation will further expand the applicability of Kafka-based event-streaming systems.

Although the review was intentionally limited to 42 studies for conciseness, the resulting taxonomy and heuristics offer practical guidance to engineers and researchers designing high-throughput, fault-tolerant data pipelines. Documenting configuration parameters and sharing benchmark artifacts are key steps toward reproducible, energy-efficient, and trustworthy event-streaming research.

\section*{Conflict of Interest}
The author declares no conflicts of interest.

\FloatBarrier


\end{document}